\documentstyle[PASJadd,epsf]{PASJ95}
\draft

\begin{document}
\setcounter{page}{1}
\setcounter{footnote}{1}

\title{X-Ray Microlensing of Bright Quasars}

\author{Rohta {\sc Takahashi},
        Atsunori {\sc Yonehara}$^1$, and
        Shin {\sc Mineshige}
\\
{\it Department of Astronomy, Graduate School of Science, Kyoto
University, Sakyo-ku, Kyoto 606-8502}
\\
{rohta@kusastro.kyoto-u.ac.jp}}
\footnotetext{Research Fellow of the Japan Society for the Promotion of Science}

\abst{
We calculated the expected microlens light curves for several 
accretion-flow models and compared the results to
find what aspects of distinct flow structures
can be extracted from the microlens light curves.
As for the flow models, we considered the standard disk, 
optically thin advection-dominated flow, and
the disk-corona model by Kawaguchi et al.,
which can nicely reproduce the observed quasar spectra.
The calculated microlens light curves of the disk-corona model exhibit
rapid soft X-ray and slow hard X-ray variations.  This is because
the former is generated by Compton up-scattering of soft (optical -- UV)
photons from the innermost part of the disk body, while the latter is
emitted via bremsstrahlung within the corona at relatively large radii. 
Furthermore, an inhomogeneous emissivity distribution over the disk
produces humps in the microlens light curves because of
enhanced microlens amplification of the bright areas.
It is hoped that black-hole accretion theory coupled with future
multi-wavelength microlens observations will be able to reveal such
spatial and temporal behaviors of flows in the vicinity of black holes.
}
\kword{accretion, accretion disks --- black holes ---
galaxies: active --- galaxies: Seyfert}

\maketitle

\section{Introduction}

Although the view of black-hole accretion as a central engine producing 
quasar activity
had already been established in the 1960's (e.g., Lynden-Bell 1969), 
our understanding of the basic flow structure 
is still in a stage far from being satisfactory.
One of the reasons is obvious: we are not able to resolve
accretion-flow structure with any existing present-day telescopes.
Eventually,  next-generation space VLBI 
and/or X-ray interferometers in the space (Cash et al. 2000) 
will be able to detect electromagnetic radiation with
superb angular resolution.
However, even without such a great innovation in technology,
there is a way to investigate the structure of quasar accretion disks 
which is a technique using microlensing (Chang, Refsdal 1979,
1984; Blandford, Hogg 1985).
It is then even possible to map the quasar disk structure 
in detail (Yonehara et al. 1998, 1999).
Broad band photometry will be able to detect any color changes, 
thereby revealing the structure of quasar accretion disks.
Here, we elucidate the theory of microlens diagnostics on quasar structure.

There exists at least one ideal source for this purpose: 
that is, Q 2237+0305, the so-called Einstein Cross (Huchra et al. 1985).  
In this system,
all four images observed in a cross shape
are very close, $\sim 1^{\prime\prime}$,
to the image center of the lensing galaxy (e.g., Irwin et al. 1989).
This situation indicates the path of photons being almost symmetric, and 
permits us to neglect any time delay between the images; 
e.g., the longest time delay, i.e., between images A and B, is only a few hours
(Wambsganss, Paczy\'nski 1994). 
Since the microlens light variation takes places on timescales of a few months,
we can easily discriminate microlensing events from intrinsic variability; 
if only one image exhibits a peculiar brightening over several tens of days 
superposed on intrinsic variations which the other three also show, 
we can conclude that the event is due to microlensing.  
Variabilities caused by supernovae etc. are distinguishable from 
microlensing in terms of their characteristic shapes of light curves.

In the present study, we focused on exploring microlens light variations
expected for luminous quasars based on the disk-corona accretion-flow model 
proposed by Kawaguchi et al. (2001).
We, here, consider the forward problem; i.e., calculate microlens light curves
for given accretion flow models.  Since we have plenty of free parameters,
such as the disk inclination and velocities of lensing stars, to be determined,
it is important to see as a first step
what aspects of the flow structure appear in the microlens light curves,
in parallel with elaborating the technique of the inverse problem
(Agol, Krolik 1999; Mineshige, Yonehara 1999).
X-ray emission from Q 2237 was detected 
by Wambsganss et al. (1999) with ROSAT in the phase without
microlensing. 
Monitoring caustic crossing will eventually be able to
reveal the enigmatic flow structure of the central region.

This paper is organized as follows.
We first describe the basic methodology of microlens diagnostics
and overview the current flow models in section 2.  
We then calculate the expected multi-wavelength microlens light curves and
discuss what they imply in section 3 and   
discussion on observational implications, with special attention placed on 
inhomogeneous flow structure, is presented in section 4.  
The final section is devoted to conclusions.

\section{Methods of Calculations}

\subsection{Einstein Cross Q 2237+0305: Basic Scales}

There are two important length scales in microlensing phenomena.
One is the Einstein-ring radius on the source plane, 
\begin{equation}
 r_{\rm E} \equiv \theta_{\rm E}D_{\rm os}
           \equiv \left(\frac{4GM_{\rm lens}}{c^2}
                 \frac{D_{\rm ls}D_{\rm os}}{D_{\rm ol}}\right)^{1/2},
\end{equation}
where $\theta_{\rm E}$ is the Einstein-ring radius;
$M_{\rm lens}$ is the typical mass of a lens star; 
and $D_{\rm ls}$, $D_{\rm os}$, and $D_{\rm ol}$ 
represent the angular diameter distances from lens to source, 
from observer to source, and from observer to lens, respectively 
(Schneider et al. 1992). For Q 2237+0305 (e.g., Huchra et al. 1985), 
the redshifts corresponding to the distances from observer 
to the quasar and from observer to the lens are
$z_{\rm os} = 1.675$ and $z_{\rm ol} = 0.039$, respectively.
The Einstein-ring radius on the source plane is then
\begin{equation}
\label{eq:E}
 r_{\rm E} \sim 1.5\times 10^{17} 
                \left(\frac{M_{\rm lens}}{M_\odot}\right)^{1/2} {\rm cm},
\end{equation}
where we assume the Einstein--de Sitter universe and
Hubble's constant to be $H_{\rm 0} \sim 60~{\rm km \ s^{-1} Mpc^{-1}}$, 
according to Kundi\'c et al. (1997).

The second, more important length scale is
a caustic crossing length over the quasar image plane per unit time,
\begin{equation}
\label{eq:cross}
 r_{\rm cross} 
    = v_{\rm t}t \frac{D_{\rm os}}{D_{\rm ol}} 
           \sim 6.9\times 10^{13}
		   v_{1000}
               \left(\frac{t}{1~{\rm d}}\right) {\rm cm},
\end{equation}
where $t$ is a certain time which the caustic moves on 
the quasar disk plane and $v_{1000} \equiv v_{\rm t}/1000$ km s$^{-1}$
with $v_{\rm t}$ being the transverse velocity of the lens object
on the lens plane
including the transverse velocity of the peculiar motion of 
the foreground galaxy relative to the source and the observer.
Fortunately, this crossing length is comparable to the Schwarzschild radius
for a $10^8M_8 M_\odot$ black hole, $r_{\rm g}\simeq 3\times 10^{13}M_8$cm,
and is much smaller than $r_{\rm E}$.  
Namely, due to a finite source-size effect, we are able to resolve the
source structure on scales much less than the Einstein-ring radius.
By daily observations we can resolve the disk structure 
with a good spatial resolution.

\subsection{Simulation Methods of Gravitational Microlensing}
Let us next explain how to
calculate light variations based on specific disk models. 
The observed flux from a part of the disk at ($r_i,\varphi_j$) 
in the cylindrical coordinates
during a caustic crossing is calculated by
\begin{equation}
\label{flux}
 F_{\rm obs}(\nu;r_i,\varphi_j) d \Omega 
  \simeq A(u) \frac{F[\nu(1+z_{\rm os});r_i]~r_i\Delta r\Delta\varphi}
                   {4\pi D_{\rm os}^2(1+z_{\rm os})^3}.
\end{equation}
Here, $A(u)$ represents the amplification factor, 
$u = u(r_i,\varphi_j,t)$ is the angular separation between a point
$(r_i,\varphi_j)$ on the source and the caustic in units of $r_{\rm E}$,
and $F(\nu;r_i)$ is the flux at frequency $\nu$ from a unit area on
a concentric ring at $r = r_i$ (axisymmetric emission is assumed;
see Yonehara et al. 1998 for more details).
Note that we take the
angular separation to be positive (or negative), $u>0$ ($u<0$),
when the source is placed inside (outside) the caustics.
The integration $\int F_{\rm obs}(\nu; r,\varphi)d\Omega$ over the disk plane 
gives the total observed flux at $\nu$.
We consider an idealized situation of a straight caustic
with infinite length passing over the disk,
since the size of the disk in question is, at most, of the order of 
several hundreds of Schwarzschild radii, $\sim 10^{16}M_8$cm,
much less than the Einstein-ring radius on the source plane 
[equation (\ref{eq:E})].  Then, the following analytical 
approximate formula can be used (see Schneider et al. 1992):
\begin{equation}
A(u) = \left\{
	\begin{array}{@{\,}ll}
       A_{\rm 0} + k u^{-1/2} & \mbox{(\rm for $u > 0$)} \\
	 A_{\rm 0} & \mbox{(\rm for $u \le 0$)}.
	\end{array}
       \right.
\label{eq:ampli}
\end{equation}
Here,  $A_{\rm 0}$, represents the initial amplification factor; i.e.,
amplification 
due to the macrolens effect, as well as the effects of other caustics. 
For the amplification factor ${\rm A}_0$ of image C 
of Q 2237+0305, we assign  ${\rm A}_0=2.5$ (Schmidt et al. 1998). 
The variation in the $V$ magnitude of the image C during the
 possible microlens event in 1999 (Yonehara 2001) is more than 
0.5 mag from the OGLE web page
\verb|(http:/www.astro.princeton.edu/~ogle/ogle2/huchra.html)|.
Therefore, we set the parameter $k=0.3$ so as to roughly reproduce 
a variation of 0.5 mag in the $V$ band around $\sim 5500$ \AA.

\subsection{Accretion Flow Models}
The emergent flux, $F(\nu; r_i)$, in equation (\ref{flux}) 
depends on the specific accretion flow model.
We here consider three representative flow models:
the standard disk, an optically thin ADAF, 
and a composite disk-corona model.

\subsubsection{Standard models}
A key assumption of the standard disk model 
(Shakura, Sunyaev 1973; Novikov, Thorne 1973)
resides in the energy balance; 
viscous heating is balanced with radiative cooling,
$Q_{\rm vis}^+ = Q_{\rm rad}^-$.
We can then build up a set of the basic equations,
from which we can uniquely determine the effective temperature
distributions, $T_{\rm eff}(r)$, for a given black-hole mass,
$M$, and mass-flow rate, $\dot M$, as
\begin{equation}
\label{Teff}
  T_{\rm eff} \simeq 2.2 \times 10^5
       {\dot M_{26}}^{1/4}M_8^{1/4}r_{14}^{-3/4}
       \left(1-\sqrt{\frac{r_{\rm in}}{r}}\right)^{1/4} {\rm K},
\end{equation}
where ${\dot M_{26}}\equiv {\dot M}/10^{26}$g s$^{-1}$,
$M_8\equiv M/10^8 M_{\odot}$, 
$r_{14}\equiv r/10^{14}$ cm; we ignored
relativistic effects, irradiation heating, 
 and Compton scattering effects.
The inner edge of the disk is set to be at $r_{\rm in} =3~r_{\rm g}$.

In the present study, 
the mass of the black hole at the center of the disk is fixed to be
$10^8 M_\odot$ and the mass-accretion rate is determined 
so as to reproduce the observed $V$ magnitudes of the
Einstein Cross in the absence of microlensing.
(Note that for such $\dot M$, the calculated X-ray luminosity 
at 0.1--2.4 keV is lower than the observed value by ROSAT by some factor.
This is probably due to significant absorption in the real case.)

\begin{figure}[t]
 \epsfxsize\columnwidth \epsfbox{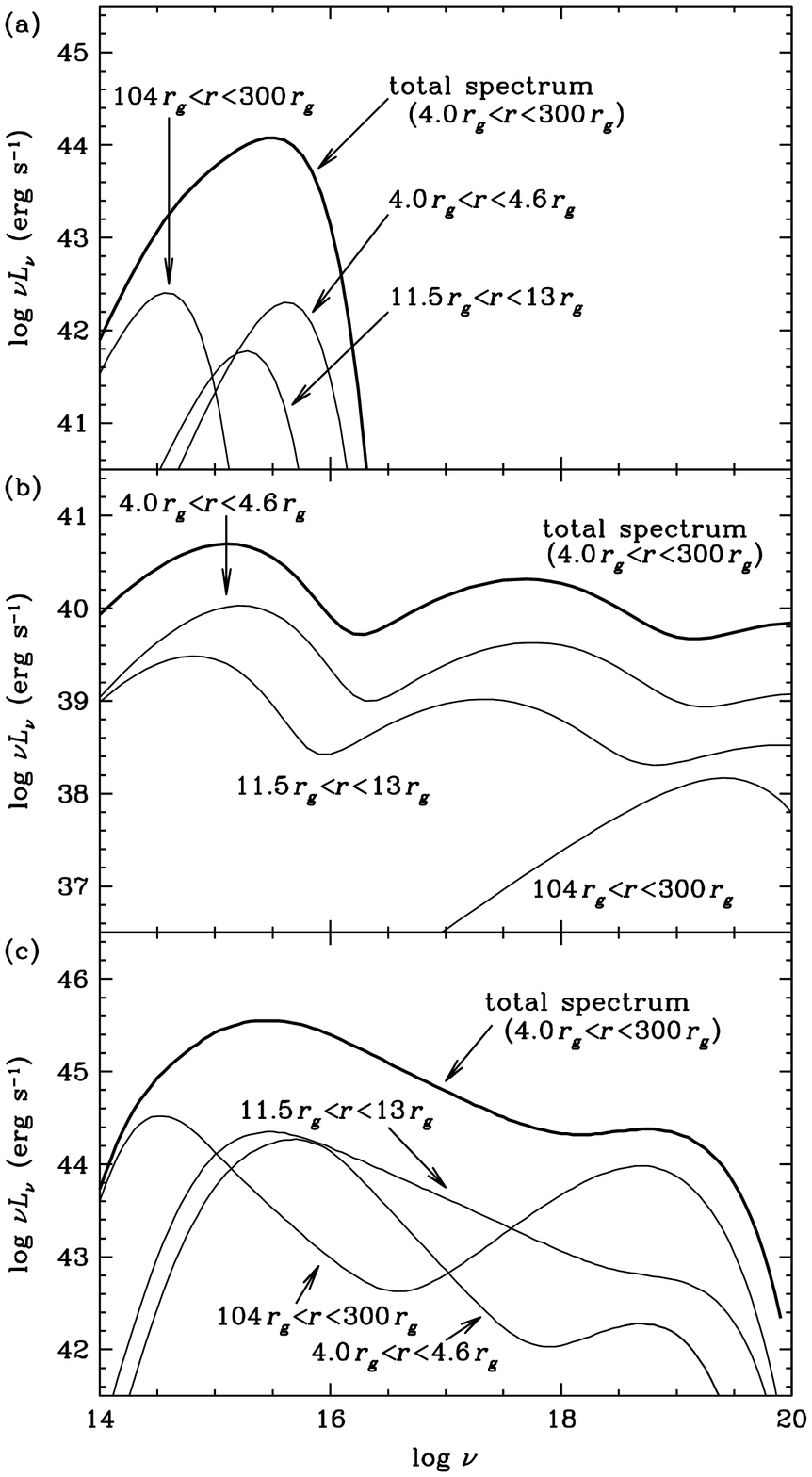}
\caption{
Spectral energy distribution in the rest frame
of a standard disk (a: top), 
an optically thin ADAF (b: middle), and
a composite disk-corona structure (c: bottom), respectively,
all with the contributions by individual concentric rings
being plotted with thin lines.}
\end{figure}

The resultant total spectrum and 
the local spectrum emitted from each concentric ring
is displayed with the thick and thin lines, respectively, in figure 1a.
Two prominent spectral features are expected for a standard disk.
One is that the disk is optically very thick, thus producing
blackbody radiation, $B_\nu(T)$, with a temperature of $T\sim 10^5$ K.
Hence, we set $F(\nu;r) = B_{\nu}\left[T(r)\right]$.
As a consequence, 
the emitted photon energy range is rather restricted to optical--UV ranges
with practically neither X-rays nor radio emissions being possible.
Such a bump in the UV regimes is quite reminiscent of the big blue bump (BBB) 
commonly observed in AGNs (Shields 1978; Malkan 1983).
The other is that the brightness distribution
simply reflects the depth of the gravitational potential well;
i.e., the disk is hotter inside and cooler outside for  
$T_{\rm eff} \propto r^{-3/4}$ [see equation (\ref{Teff})].  Therefore,
we expect short-wavelength radiation to come from the inner compact
region, while long-wavelength radiation would originate from rather 
large areas.

\subsubsection{ADAF: Advection-Dominated Accretion Flow}

The next interesting solution is an
optically-thin advection-dominated accretion flow (ADAF;
Ichimaru 1977; Narayan, Yi 1995; Abramowicz et al. 1995).
The key relation is again in the energy balance, which is
\begin{equation}
  Q_{\rm adv} \equiv \Sigma T v_r\frac{\partial s}{\partial r}
      = Q_{\rm vis}^+ \gg Q_{\rm rad}^-,
\end{equation}
where $\Sigma=\int\rho dz$ is the surface density,
$s$ the specific entropy and $v_r$ the accretion velocity
(see Kato et al. 1998 for the detailed discussion).
The term on the left-hand side represents the advective energy
transport, i.e., heat transport carried by accreting gas.

In contrast to the standard disk, in which 
the accretion energy efficiently goes into radiation energy, 
the accretion energy of gas in ADAF turns into its internal energy, 
with a small fraction being radiated (Ichimaru 1977).  
Because of an entirely different energy balance,
the optically thin ADAFs are, in many respects, 
distinct from the standard Shakura--Sunyaev disks.
In other words, an ADAF is a faint hot flow, 
in contrast to a cool bright disk in the standard case.
Accreting matter falls into a central object without losing
its internal energy via radiation. 
Hence, the disk can be significantly hotter,
thereby producing high--energy (X--$\gamma$ rays) photons.
Typically, ion temperatures are near virial,
whereas electron temperature is significantly lower
because of inefficient Coulomb coupling between electrons and ions, 
\begin{equation}
  T_{\rm ion} \simeq 10^{12}(r/r_{\rm g})^{-1}~{\rm K}, \quad
  T_{\rm elec} \simeq 10^{9-10}~{\rm K}.
\end{equation}
Hence, the radiation emissivity does not always reflect the shape of
the potential well. 

In figure 1b we display the spectral energy distribution
of a typical ADAF, calculated by Manmoto et al. (1997), 
to be compared with other models.
Note that the total flow luminosity is not adjusted to 
that of the Einstein Cross in this model, since
no ADAF solutions exist at high luminosities comparable to
the Eddington luminosity.

There are two unique spectral features of optically thin ADAFs.
First, ADAFs can produce broad-band spectra because of many radiation
processes being involved.
Included are synchrotron and bremsstrahlung radiation as well as  inverse
Comptonization of soft photons created by the former two processes
(see Narayan, Yi 1995 for details).
The photon energy is thus spread over a large frequency range:  
from radio (via synchrotron) to hard X--$\gamma$ rays 
(via inverse Compton effect).
Second, the radiation energy does not always reflect 
the depth of the potential well,
where photons are emitted; e.g.,
radio emission via synchrotron predominantly originates in the inner parts,
while bremsstrahlung X-ray photons are from a rather wide spatial range.
Figure 2b shows that emission from a region within 
a few tens to hundreds of $r_{\rm g}$ 
from the center contributes most of the entire spectrum.

\subsubsection{Composite disk-corona structure}

Since the optically thin ADAFs are too faint to explain 
the luminosity of bright quasars, such as Q 2237+0305,
we need an alternative model 
for making high-energy emission possibles from luminous AGNs.
It is widely believed that high-luminosity AGNs have
a composite disk-corona structure.
This idea is supported by the fact that AGNs generally
exhibit two (or more) spectral components: 
a power-law component extending up to $\sim 100$ keV
and the big blue bump at UV ranges, which is, if blackbody,
characterized by temperatures of $\sim 10^5$ K.
The coexistence of power-law and blackbody-like components is 
characteristic of spectra of typical Seyfert 1 type AGNs, which
actually indicates a composite flow structure, such as a two-phased
disk-corona structure.  
We, next, consider a hybrid disk model as a central engine in the quasar.

Although discussions concerning constructing disk-corona models 
have a long history, 
we do not yet have a sort of {\lq}standard model{\rq}, 
owing to difficulties arising from complex thermal, radiative,
 and dynamical interactions between the disk and coronal components.
 Here, we use a model by Kawaguchi et al. (2000),
since they could, for the first time,
reproduce the observed broad-band spectral properties of quasars, 
as represented by the composite spectrum 
(Zheng et al.\ 1997; Laor et al.\ 1997).
The idea is to couple a standard-type disk body with
$T_{\rm eff} \sim 10^5 (r/r_{\rm g})^{-3/4}$ K and
an advection-dominated disk corona, in which $T_{\rm elec} \sim 10^9$ K.
Following Haardt and Maraschi (1991), they assume that
a fraction, $1-f$, of energy is dissipated in the disk body,
while the remaining fraction, $f$, is in the coronae.  Here,
$f$ is fixed to be 0.6.
By solving the hydrostatic balance and 1D radiative transfer
including inverse-Compton processes self-consistently,
they finally obtain the
spectrum shown by the thick line of figure 1c,
which gives a reasonable fit to the composite spectrum.

According to this model,
the big blue bump is caused by thermal emission 
from the disk body at small radii,
the soft X-ray excess is inverse-Compton scattering of the
BBB soft photons in the corona at small radii, 
and the hard X-rays are bremsstrahlung radiation
from the coronae at large radii.
We also show the contributions of the individual rings to the total spectrum 
by thin lines in figure 1c.
Importantly, soft X-rays are from the vicinity of the black hole,
as expected, and as is in the case of ADAF,
while hard X-rays are from a rather wide areas.
This is because the temperature profile is nearly flat, but the
density goes roughly as $r^{-1}$ in the corona, leading to the 
bremsstrahlung luminosity from volume $(V \propto r^3)$
being proportional to $\rho^2 T^{1/2}V \propto r$.
Such unique emission properties produce interesting features in 
the multi-wavelength microlens light curves.

\section{Microlens Diagnostics of Accretion Flow Structure}
\subsection{Microlens Light Curves}

\begin{figure}[t]
 \epsfxsize\columnwidth \epsfbox{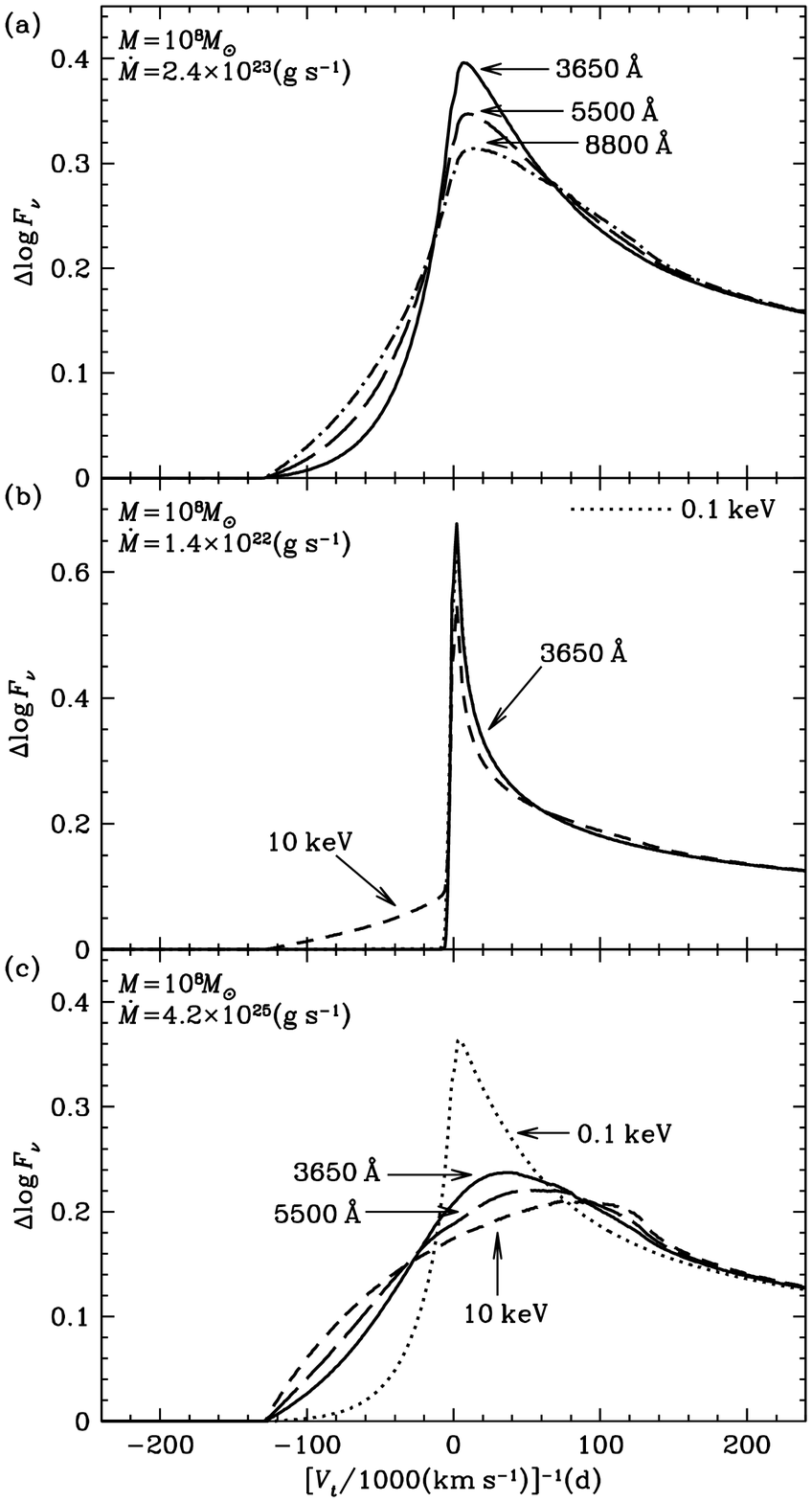}
\caption{Caustic crossing light curves of the standard disk (a: top),
the optically thin ADAF (b: middle), 
and the disk-corona model (c: bottom), respectively, at several 
energy bands(3650 \AA, 5500 \AA, 8800 \AA, 0.1 keV, and 10 keV).  
We set $\Delta F_\nu= 0$ at each $\nu$ outside the caustic.
}
\end{figure}  

We are now ready to calculate microlens light curves
of various types of accretion flows.
We focus our discussion on the relative changes in the radiation flux and 
the variation timescale, although we might note that the latter
sensitively depends on unknown parameters, such as $v_{\rm t}$
and disk inclination angles.
For simplicity, we assume that the accretion disk 
is face-on to the observer ($i=0$).
If the disk is tilted by an angle, $i~(\ne 0)$,
the absolute luminosity of the source is reduced by $\cos i$
and the variation timescale will be shortened by $\cos i$,
if the tilt is in the direction parallel to the motion of the caustic.
Conversely, 
a tilt in the perpendicular direction induces no changes in timescale.
In any case, no large alternation in the basic properties will arise.
The inclusion of relativistic effects 
(e.g., Doppler shifts by the disk rotation, 
beaming, gravitational redshifts)  
in the disk model could reveal the detailed disk structure 
in the real vicinity of a central black hole, which will be discussed 
in a future paper.

\begin{figure}[t]
 \epsfxsize\columnwidth \epsfbox{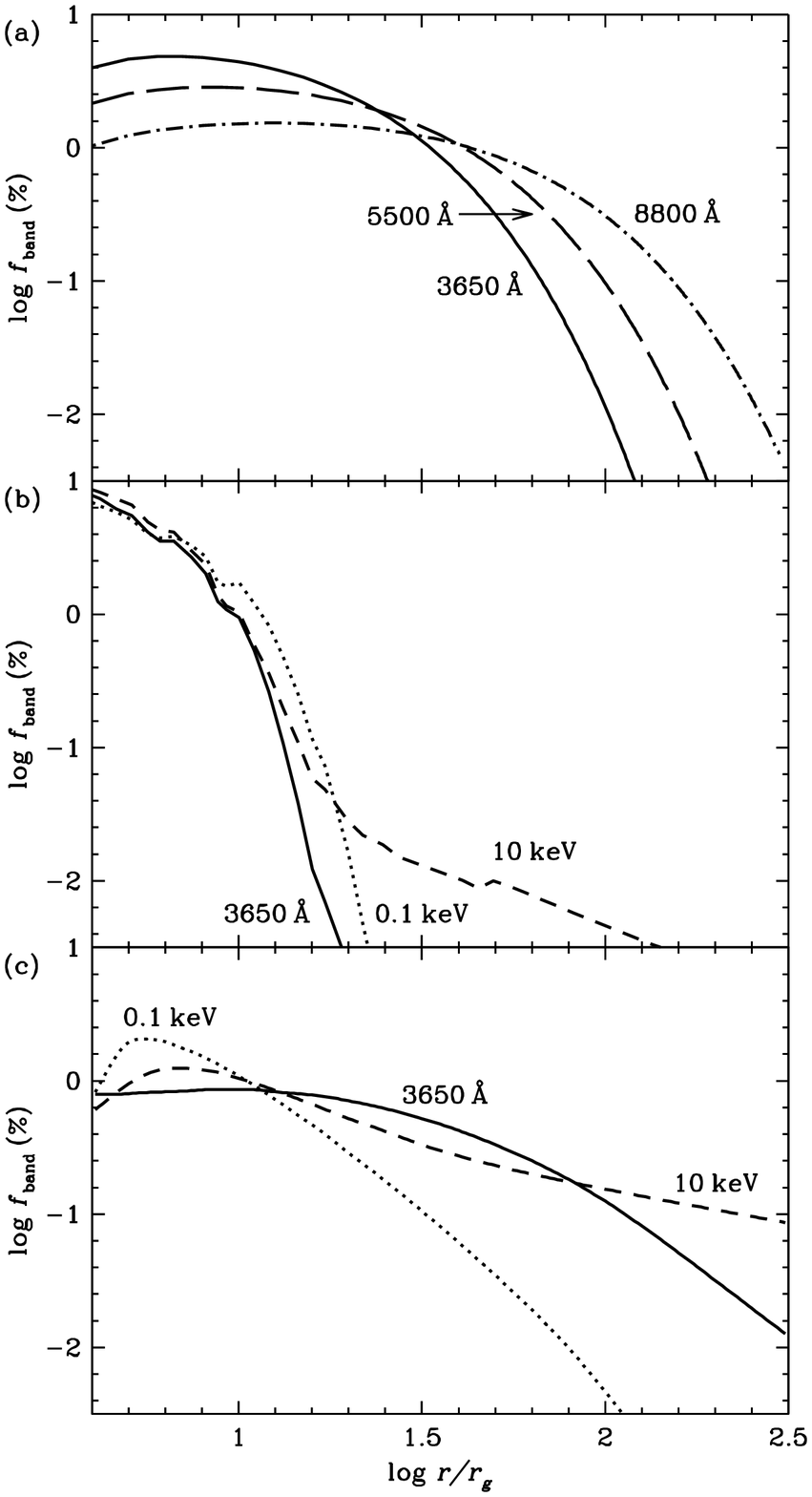}
\caption{Radial profile of radiation luminosities at several wave-bands
of the standard disk (a: top), 
the optically thin ADAF (b: middle), 
and the composite disk corona structure (c: bottom), respectively.
The ordinate is $f_{\rm band}$(\%)$\,=100\times\nu F_{\nu}\Delta r/\int\nu F_{\nu}dr$.
}
\end{figure}   

The basic tendencies of the standard and ADAF cases were
already reported by Yonehara et al. (1998).
In the case of the standard disk illustrated in figure 2a,
short-wavelength variations are more rapid than 
those of long-wavelength ones.
Obviously, microlens light curves reflect 
radial changes of the emitting photon spectra.  
To help understand the basic trends of the microlens light curves,
we plot in figure 3a the radiation spectra of the concentric rings 
in a standard disk
as a function of radius, $r$, for several wave-bands.
The inner hot, outer cool disk structure of the standard disk yields
a smaller effective size of the region emitting short-wavelength radiation 
than otherwise.  This explains
more rapid variations in short-wavelength radiation in the optical--UV bands 
(see figure 2a).

In the case of ADAF, in contrast,
both the optical and soft X-ray fluxes displayed in 
figure 2b show rather abrupt changes,
compared with optical variations of a standard-type disk.  
In other words, the emissivity does not reflect the shape of the
gravitational potential well, which goes as $1/r$.
The reason is that large magnetic-field 
and electron energy densities are achieved only in a compact region around
the black hole; thus, efficient synchrotron radiation in the radio
region is possible there,
which is Compton up-scattered to produce optical radiation and X-rays.
The microlens light curves in figure 2 catch the two aspects 
of the radial spectral energy distribution displayed in figure 3. 
The timescale of the light variations depends on the radial extent
of the emitting region, while the peak amplification of the light curves 
corresponds to the ratio of the peak emissivity to
the average value.

Finally, we represent the predicted light curves of the
disk corona model at three observed bands 
(3650 \AA, 0.1 keV, 10 keV in figure 3c).
As expected from figure 1c,
soft and hard X-ray light curves are distinct;
soft X-ray radiation shows a sharp peak around the caustic crossing time,
while hard X-ray variations are somewhat smoother.
This reflects that soft X-rays originate from the inner parts 
of the corona via Comptonization of soft photons from the disk body
and hard X-rays are from the outer parts of the corona 
emitted by bremsstrahlung.

Figure 3c clearly illustrates different emissivity distributions 
of the disk at different wavebands.
We find similar tendencies in the X-ray emissivity of the ADAF
and the disk corona system in the sense that 
soft X-ray emissivity exceeds the hard X-ray one at large radii.
However, there is a quantitative difference: 
the 0.1 keV flux exceeds 10 keV flux at $r \geq 80~r_{\rm g}$ in
the disk-corona case and at $r \geq 30~r_{\rm g}$ in the ADAF.
Such different behavior at different X-ray energy bands can be
tested by observations of microlensing.
Optical variations are also similar to those of the standard disk, which
is not surprising because the disk body emitting optical--UVs
is of a standard-disk type, although the emissivity profile
is smoother in figure 3c than in figure 3a.
This is because Compton scattering enhances
the optical flux in the disk-corona system.
We can thus infer the radial disk structure
from light curves at different frequencies.
The present calculation does not include synchrotron emission,
but if included we expect substantial radio emissions
whose variations will be sharper around the peak (see figure 2b).




\subsection{Microlens of Intrinsic Variability}
The most impressive feature of radiation from AGNs
is its rapid variability (Ulrich et al. 1997; Kawaguchi, Mineshige 1998).
Probably, such variation is not due to any coherent brightness variation
over the disk surface, but is caused by local,
transient brightening, or flares 
(Wheeler 1977; Takahara 1979; Galeev et al. 1979).
We then expect a rather inhomogeneous disk structure, such as
blobs, spots, or filaments, if hot accretion flow
is viewed by X-rays from its vicinity,
just like the Yohkoh soft X-ray image of the Sun.
Recently performed three-dimensional (3D)
magnetohydrodynamical (MHD) simulations also support this view
(Machida et al. 2000: Kawaguchi et al. 2000).
If so, such an inhomogeneous structure should inevitably affect the
microlens light curves, especially in X-rays.
Of course, un-microlensed images, themselves, suffer 
from intrinsic variabilities, but an microlensed image 
should exhibit more dramatic changes when a caustic crosses
a flare, since transient brightening occurs in much more compact regions 
than the region emitting persistent radiation.

\begin{figure}[t]
 \epsfxsize\columnwidth \epsfbox{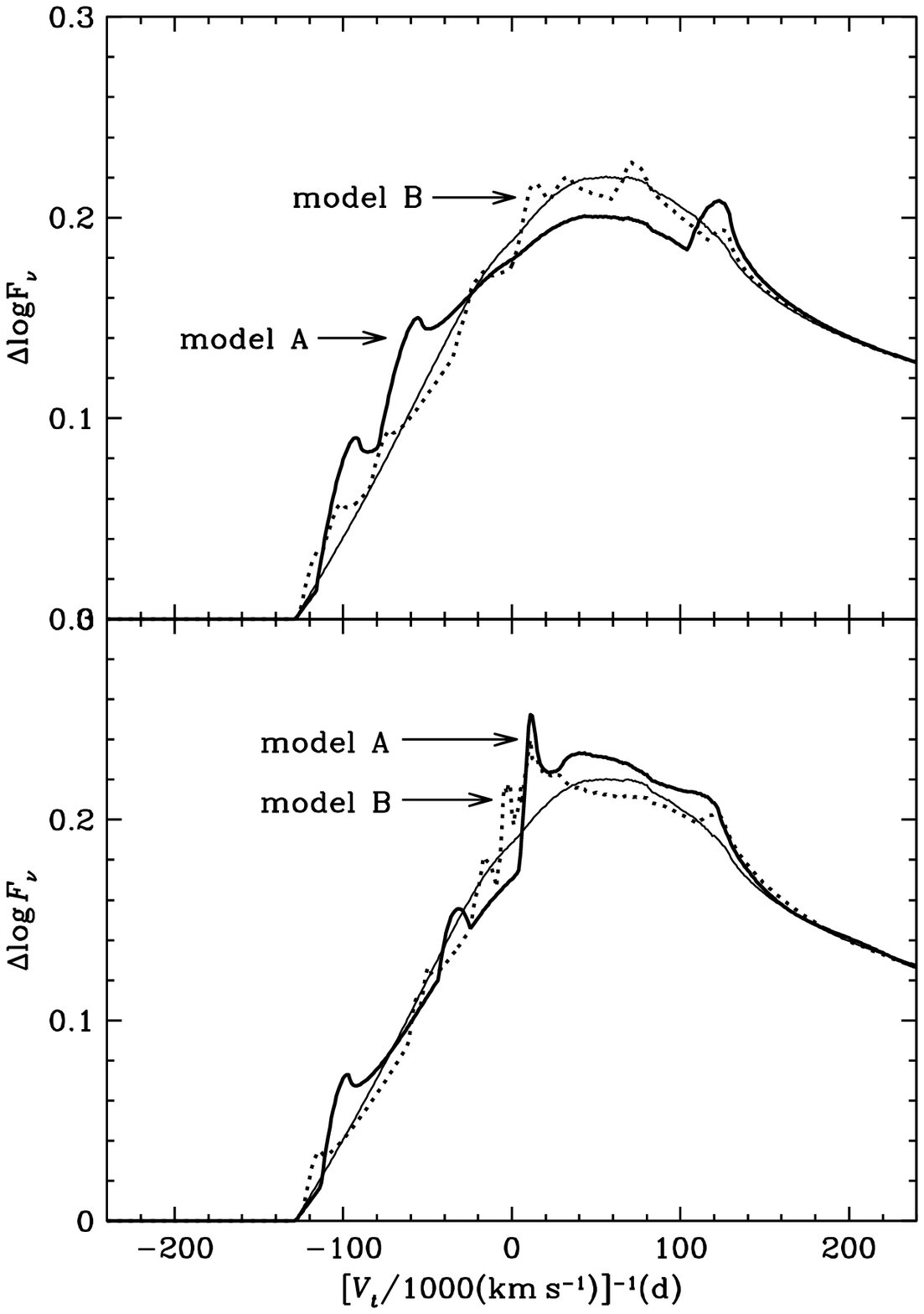}
\caption{Microlens light curves of a disk with multiple blobs 
whose positions are fixed (top) or 
moving with Keplerian rotation (bottom).
In each panel 
the thin solid line, the thick solid line and the dotted line 
represents the case with no blobs
(i.e. a uniform disk), 3 blobs (model A), and 12 blobs (model B),
respectively.}

\end{figure}   

To demonstrate such effects, we performed simple calculations;
we put 3 (model A) or 12 blobs (model B), whose total luminosity is 20\% of 
the total disk luminosity, on a disk with a flat brightness distribution
emitting the remaining fraction, 80\%, of luminosity.
We assumed a uniform disk with constant brightness, for simplicity,
and took the disk radius to be 300 $r_{\rm g}$,
while the blob radius was either $30~r_{\rm g}$ (model A)
or $15~r_{\rm g}$ (model B).
We also assumed that luminosities of the blobs were not varying in time and 
that positions of the blobs were fixed or changed 
in time according to the Keplerian rotation.  

The resultant light variations are plotted in figure 4 in comparison with
the case with no blobs (by the dashed line).
Although the total blob luminosity is only 20\% of the total,
microlens light variations record dramatic changes;
if the number of blobs is not very large, say, less than 10,
we can even count the number of blobs.
According to MHD simulations, however,
the number of blobs obeys a power-law relation against their sizes
(Kawaguchi et al. 2000); that is, there are numerous small blobs
as well as a small number of large blobs.
However, if there are big blobs, we can identify their contributions
in the microlens light curves, provided that frequent monitoring
observations (on $\sim$ d) are made.
In reality, the luminosities of blobs are likely to be time-dependent.
The duration of each hump around the disk center (at $r< 10^2r_{\rm g}$) 
is shorter when blobs are rotating.
This is because the Keplarian rotation periods,
\begin{equation}
\tau_{\rm Kep}\sim\sqrt{\frac{r^3}{GM}}\sim15
\left(\frac{r}{10^2r_{\rm g}}\right)^{3/2}\left(\frac{M}{10^8M_\odot}\right)~{\rm d},
\end{equation}
are shorter than the caustic crossing timescale of blobs, 
\begin{equation}
\tau_{\rm blob}\sim20
\left(\frac{\ell_{\rm blob}}{60~r_{\rm g}}\right)
\left(\frac{v_{\rm t}}{1000~{\rm km~s^{-1}}}\right)^{-1}~{\rm d},
\end{equation}
where $\ell_{\rm blob}$ is the blob diameter and $v_{\rm t}$ is the 
transverse velocity of the caustics. 

\section{Discussion} 

We calculated the expected microlens light curves for several disk
models and compared the results to
find what aspects of distinct flow structures
can be extracted from the microlens light curves.
Neither a standard-type disk nor optically thin ADAF
is adequate to explain the emissions from bright quasars, since 
the former cannot emit substantial X-rays and
the latter solution is found only in low-luminosity regimes.
Intense X-ray emission is expected only from a composite disk-corona
structure, although its detailed structure is unknown, unfortunately.
We used a recent model by Kawaguchi et al. (2000) to see
basic tendencies of multiwavelength light curves.
What has become clear is that,
if soft X-rays are generated by Compton up-scattering of 
soft (optical--UV) photons emerging from the disk body
and if hard X-rays are due to bremsstrahlung, 
we can expect a more rapid soft X-ray variation than hard X-ray changes.

There have been interesting discussions on the constraints
of the source size by using the microlens in Q 2237+0305.
An upper limit to the size of the emitting region
set by Wambsganss et al. (1990) and by Rauch and Blandford (1991)
is $2\times 10^{15}$ cm at $1\mu$m, corresponding to
$\log\nu \sim 15.0$ in the quasar rest frame.
By constructing a simple disk model which mimics the standard disk,
Rauch and Blandford (1991) concluded 
that the standard-type accretion disk model can be ruled out.
Jaroszy\'nski et al. (1992), conversely, assert
that the standard disk model is consistent with the observations,
if we adopt the realistic disk model.
Czerny et al. (1994) performed more extended, thorough investigations,
reaching the conclusions as follows:
(1) a standard disk model is marginally consistent with the data,
(2) an irradiated disk model can both meet the size requirements
and reproduce the slope of the quasar spectrum, and
(3) even a model with a huge number of small optically thin cloudlets
is marginally consistent with the data.
Back to our computations, figure 3 shows that all of our models
can meet the source-size constraint of $\lsim 2 \times 10^{15}$ cm,
corresponding to $70 M_8 r_{\rm g}$ at optical bands.

We also examined how inhomogeneity in disk brightness profile affects 
microlens light curves. We expect significant peaks superposed on 
gradual microlens variations.
Further, the detailed shapes of the peaks are subject to the motion of
blobs. 
Hence, the microlens light curves suffer both from changes in
the light amplification pattern due to caustic crossing
and time variations of blob properties.
It will not always be easy to decouple these two effects.
In any case, we may be able to obtain information regarding the site and 
emission processes of the intrinsic variability from frequently sampled
microlens light curves at various wavelengths in the near future.

We wish to stress that the optical emission properties cannot
impose strong constraints on the flow model, 
since low-energy photons can be created by a number of mechanisms 
which can work, in principle, at any radii.
That is, X-ray observations are indispensable 
to investigate the vicinity of the black hole in Q 2237+0305.
The first X-ray observations were made by Wambsganss et al. (1999) 
with ROSAT in the energy band of 0.1--2.4 keV.
Because of the maximum image separation of only 1$^{\prime\prime}$.8,
the ROSAT/HRI detector was unable to resolve the four images 
of Q 2237+0305.  
Chandra X-ray Observatory with a much better on-axis resolution 
of 0$^{\prime\prime}$.5  
can resolve the four images of Q 2237+0305 
(G. Chartas, 2000 private communication).
More important are the
simultaneous multi-wavelength observations during a microlens event.
For this purpose, Chandra (P.I. S. Mineshige), HST (P.I. R. Webster), 
and ground-based optical telescopes (OGLE, APO, etc)
are planned to be used; they will
be able to determine the multi-wavelength emission properties of
the quasar, thereby
testing the basic radiation processes in the disk corona structure.

\section{Conclusions}
\begin{enumerate}
\item Microlens light curves contain fruitful information regarding
the radial accretion flow structure (including magnetic-field
strength) as well as basic emission processes.

\item We expect distinct behavior in the soft and hard X-ray
microlens variations of bright quasars.  
This is because
soft X-ray emission is due to Compton up-scattering of soft (optical -- UV)
photons from the innermost part of the disk body,
and hard X-ray radiation is via bremsstrahlung within the corona
at relatively large radii.  Accordingly, soft X-rays will show
rather gradual variations, compared with hard X-rays.

\item If the AGN intrinsic variability is due to multiple flares occurring
occasionally here and there, we expect humps in the overall
microlens light curves.  Future observations will reveal
any inhomogeneity of disk emission, and may pose important
constraints on the emission processes.
\end{enumerate}
We are grateful to an anonymous referee for a useful discussion
regarding rotating blobs.
This work was supported in part
by Research Fellowships of the Japan Society for the
Promotion of Science for Young Scientists, 9852 (AY),
and by the Grants-in Aid of the
Ministry of Education, Culture, Sports, Science and Technology of Japan
(10640228, SM).

\section*{References}
\re Abramowicz, M.A., Chen, X., Kato, S., Lasota, J.-P., \& Regev, O.  1995,
   ApJ, 438, L37
\re Agol, E., \& Krolik, J. 1999, ApJ, 524, 49
\re Blandford, R.D., \& Hogg, D.W. 1995, in IAU Symp. 173,
    in Astrophysical Application of Gravitational Lensing,
     ed. C.S. Kochanek, \& J.N. Hewitt (Dordrecht, Kluwer), 355
\re Cash, W., Shipley, A., Osterman, S., \& Joy, M. 2000, Nature, 407, 160
\re Chang, K., \& Refsdal, S. 1979, Nature, 282 561
\re Chang, K., \& Refsdal, S. 1984, A\&A, 132, 168
\re Czerny, B., Jaroszy\'nski, M., \& Czerny, M. 1994, MNRAS, 268, 135
\re Galeev, A.A., Rosner, R., \& Vaiana, G.S. 1979, ApJ, 229, 318,
\re Haardt, F., \& Maraschi, L. 1991, ApJ, 380, L51
\re Huchra, J., Gorenstein, M., Kent, S., 
    Shapiro, I. I., Smith, G. Horine, E., \& Perley, R.,1985, AJ, 90, 691
\re Ichimaru, S. 1977, ApJ, 214, 840
\re Irwin, M.J., Webster, R.L., Hewett, P.C., Corrigan, R.T., \& 
    Jedrzejewski, R.I. 1989, AJ 98, 1989
\re Jaroszy\'nski, M., Wambsganss, J., \&   Paczy\'nski, B. 1992,
    ApJ 396, 65L
\re Kato, S., Fukue, J., \&  Mineshige, S. 1998,
    Black-Hole Accretion Disks (Kyoto, Kyoto University Press) p.271
\re Kawaguchi, T., \&  Mineshige, S. 1998,
    in Active Galactic Nuclei and Related Phenomena,
    ed. Y. Terzian, D. Weedman, and E. Khachikian,
    (ASP Conf. Ser. 194, Dordrecht), p356
\re Kawaguchi, T., Mineshige, S., Machida, M., Matsumoto,  \& R., Shibata, K.
    2000, PASJ 52, L1
\re Kawaguchi, T., Shimura, T., \&  Mineshige, S. 2001, ApJ, 546, 966
\re Kundi\'c, T., Turner, E. L., Colley, W. N., Gott, J. R. III, 
    Rhoads, J. E., \& Gloria, K. A. 1997, ApJ 482, 75
\re Laor, A., Fiore, F., Elvis, M., Wikes, B. J., \&  McDowell, J. C. 1997,
    ApJ, 477, 93
\re Lynden-Bell, D. 1969, Nature, 223, 690
\re Machida, M., Hayashi, M., \&  Matsumoto, R. 2000, ApJ, 532, L67
\re Malkan, M.A. 1983, ApJ, 268, 582
\re Manmoto, T., Mineshige, S., \&  Kusunose, M. 1997, ApJ, 489, 791
\re Mineshige, S., \&  Yonehara, A. 1999, PASJ, 41, 497
\re Narayan, R.,  \& Yi, I. 1995, ApJ, 452, 710
\re Novikov, I.D., Thorne, K.S. 1973, in Black Holes,
    ed. C. DeWitt  \&  B. S. DeWitt (New York, Gordon and Breach) p343
\re \O stensen R., et al. 1996, A\&A 309, 59
\re Rauch, K.P., \&  Blandford, R.D. 1991, ApJ, 381, L39
\re Schmidt, R., Webster, R.L., \&  Lewis, G.F. 1998, MNRAS, 295, 488
\re Schneider, P., Ehlers, \&  J., Falco, E.E. 1992,
      Gravitational Lenses (Berlin: Springer-Verlag)
\re Shakura, N.I., \&  Sunyaev, R.A. 1973, A\&A 24, 337
\re Shields, G.A. 1978, Nature, 272, 706
\re Takahara, F. 1979, Progr. Theor. Phys., 62, 629
\re Ulrich, M.-H., Maraschi, L., \&  Urry, C.M. 1997, ARA\&A, 35, 445
\re Wambsganss, J., Brunner, H., Schindler, \& S. Falco, E. 1999, A\&A, 346, L5
\re Wambsganss, J.,  \& Paczy\'nski, B. 1994, AJ, 108, 1156
\re Wambsganss, J., Paczy\'nski, B.,  \& Katz, N. 1990, ApJ, 352, 407
\re Wheeler, J.C. 1977, ApJ, 214, 560
\re Yonehara, A. 2000, ApJ, 548, 127
\re Yonehara, A., Mineshige, S., Fukue, J., Umemura, M., \& Turner, E.L.
     1999, A\&A, 343, 41
\re Yonehara, A., Mineshige, S., Manmoto, T., Fukue, J., Umemura, M.,
     \& Turner, E.L.  1998, ApJ, 501, L41; Erratum 511, L65
\re Zheng, W. Kriss, G.A., Telfer, R.C., Grimes, J.P., \& 
    Davidsen, A.F. 1997, ApJ, 475, 469

\label{last}

\end{document}